\documentclass[%
 reprint,
 amsmath,amssymb,
 aps,
]{revtex4-2}

\usepackage{graphicx}
\usepackage{dcolumn}
\usepackage{bm}

\begin{document}

\preprint{APS/123-QED}

\title{Differentiable Preisach Modeling for Characterization and Optimization of Accelerator Systems with Hysteresis}

\author{R. Roussel}
\email{roussel@slac.stanford.edu}
\author{A. Edelen}
\author{D. Ratner}
\affiliation{SLAC National Accelerator Laboratory, Menlo Park, CA 94025, USA}
\author{K. Dubey}
\author{J.P. Gonzalez-Aguilera}
\author{Y. K. Kim}
\affiliation{University of Chicago, Illinois 60637, USA}
\author{N. Kuklev}
\affiliation{Advanced Photon Source, Argonne, Illinois 60439, USA}

\date{\today}

\begin{abstract}
Future improvements in particle accelerator performance is predicated on increasingly accurate online modeling of accelerators. 
Hysteresis effects in magnetic, mechanical, and material components of accelerators are often neglected in online accelerator models used to inform control algorithms, even though reproducibility errors from systems exhibiting hysteresis are not negligible in high precision accelerators. 
In this work, we combine the classical Preisach model of hysteresis with machine learning techniques to efficiently create non-parametric, high-fidelity models of arbitrary systems exhibiting hysteresis. 
We demonstrate that our technique accurately predicts hysteresis effects in physical accelerator magnets.
We also experimentally demonstrate how these methods can be used in-situ, where the hysteresis model is combined with a Bayesian statistical model of the beam response, allowing characterization of hysteresis in accelerator magnets solely from measurements of the beam. 
Furthermore, we explore how using these joint hysteresis-beam models allows us to overcome optimization performance limitations when hysteresis effects are ignored. 
\end{abstract}

\maketitle

Hysteresis is a well-known physical phenomenon where the state of a given system is dependant on its historical path through state-space.
This property is evident in physical, biological, chemical and engineering processes, including the magnetization of ferromagnetic materials \cite{jiles_theory_1986}, the activation of embryonic cells \cite{pomerening_building_2003}, the charging and discharging cycles of nickel-metal hydride batteries \cite{sauer_batteries_2009} and the driving of mechanical actuators with backlash \cite{warnecke_backlash_2003}.
In particular, hysteresis effects in magnetic \cite{sammut_measurement_2008}, mechanical \cite{huque_accelerated_2015} and material \cite{turner_no_2022} elements of particle accelerators makes optimizing the performance of current accelerator facilities used for scientific discovery challenging.

Model based optimization algorithms, such as Bayesian optimization (BO) \cite{snoek_practical_2012}, use online computational models to tackle these optimization tasks at accelerator facilities \cite{duris_bayesian_2020, roussel_turn-key_2021-1, kirschner_adaptive_2019}.
However, models used in these algorithms ignore hysteresis effects entirely, degrading optimization performance due to errors caused by hysteresis \cite{hanuka_demonstration_2021}.
This problem is expected to worsen, as ambitious targets for future accelerator performance \cite{borland_upgrade_2018, aicheler_multi-tev_2012} become increasingly sensitive to hysteresis effects.
Incorporating an accurate description of hysteresis into models used for online accelerator optimization could substantially improve the performance of current and future particle accelerators.

\begin{figure*}[ht]
    \includegraphics[width=\linewidth]{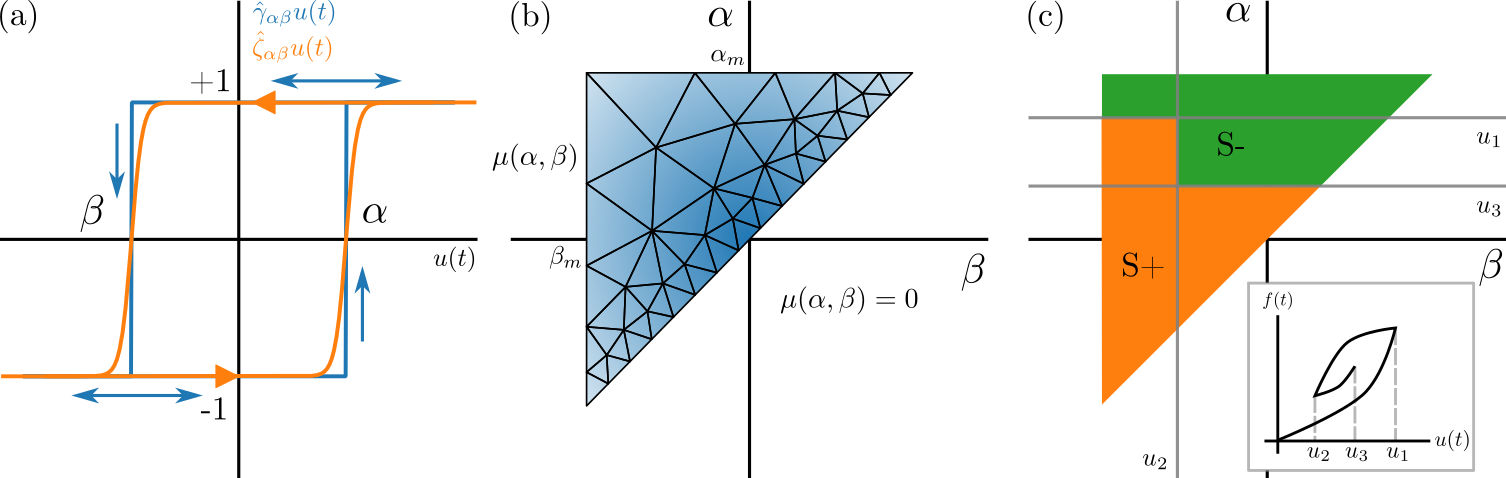}
    \caption{(Color online) Elements of the differentiable non-parametric Preisach hysteresis model. (a) Output of the hysteron operator $\hat{\gamma}_{\alpha\beta}$ and the approximate differentiable hysteron operator $\hat{\zeta}_{\alpha\beta}$ acting on the input $u(t)$.  (b) Discretization of the density on the Preisach ($\alpha$-$\beta$) plane. Note that $\mu(\alpha,\beta) = 0$ if $\alpha<\beta$, $\alpha > \alpha_m$ or $\beta < \beta_m$ where $\alpha_m,\beta_m$ are equal to the maximum and minimum inputs of the model respectively. (c) $S^+$ and $S^-$ sub-domains after three time steps, where $u_1>u_3>u_2>\beta_m$, assuming that all hysterons are in the negative state initially. Inset: Corresponding model output (not to scale). }
   \label{fig:cartoon}
\end{figure*}

Non-parametric Preisach modeling \cite{hoffmann_least_1989} is a flexible approach for accurately describing systems that exhibit hysteresis behavior.
Unfortunately, fitting these models to experimental measurements using numerical optimization techniques has been shown to be computationally expensive \cite{iyer_hysteresis_2004, marouani_implementation_2019} due to the large number of free model parameters and the so-called ``curse of dimensionality" \cite{bellman_adaptive_1961}.

In this work, we construct a \emph{differentiable}, non-parametric Preisach model, which when used in conjunction with gradient based optimization, significantly reduces the computational cost of model identification.
We explore how our technique enables accurate online modeling of the beam response with respect to controllable accelerator parameters through the use of joint hysteresis-Bayesian statistical models.
We experimentally demonstrate how this enables the characterization of hysteresis properties in magnetic beamline elements from beam-based measurements.
Finally, we explore how the joint model improves optimization of a realistic beamline containing magnetic elements exhibiting hysteresis.

The Preisach model of hysteresis \cite{mayergoyz_generalized_1988, bertotti_science_2006} is comprised of a continuous set of \emph{hysterons}, which when added together, model the output of a hysteretic system $f(t)$ for a time dependant input $u(t)$.
Given a set of discrete time ordered inputs $u_i = u(t_i)$, the hysteron state is represented by the hysteron operator shown in Fig. \ref{fig:cartoon}a $\hat{\gamma}_{\alpha\beta}$, which has an output of $\pm1$, where $\alpha$ and $\beta$ describe the input required to switch the hysteron between its two possible states. 
The number of hysterons with values ($\alpha,\beta$) is given by the hysteron density function $\mu(\alpha,\beta)$, plotted on the Preisach ($\alpha$\nobreakdash-$\beta$) plane (Fig. \ref{fig:cartoon}b).

The Preisach model output is represented by
\begin{equation}
    f(t) = \hat{\Gamma}u(t) = \iint_{\alpha\geq\beta}\mu(\alpha, \beta)\hat{\gamma}_{\alpha\beta}u(t) d\alpha d\beta
    \label{eqn:preisach_model}
\end{equation}
where $\alpha\geq\beta$ results from physical conditions of the hysteron operator.
This integral is evaluated through a geometric interpretation, shown in Fig. \ref{fig:cartoon}(c).
Given the sequence of input values $u_i$, we can determine sub-regions of the Preisach plane, $S^+$ and $S^-$, where hysteron operators output positive and negative states respectively.
We start with the assumption that all hysterons are initially in the negative state ($S^-$ covers the entire Preisach plane).
When $u_t>u_{t-1}$, a horizontal line is swept up, flipping hysteron states from negative to positive, increasing the $S^+$ region. Conversely, when $u_t<u_{t-1}$, a vertical line is swept to the left, flipping states from positive to negative.

Once the regions $S^+$ and $S^-$ are determined by the input $u(t)$, fitting a Preisach model to experimental data requires the determination of the hysteron density function $\mu(\alpha, \beta)$, often referred to as the \emph{identification problem}.
Approaches for solving this problem are generally divided into parametric or non-parametric methods.
Parametric methods describe the hysteron density using one of several analytic functions with a small number of free parameters \cite{sutor_preisach-based_2010,hergli_identification_2019,marouani_implementation_2019}, which can be determined through numerical optimization methods given experimental data.
However, this limits model flexibility, resulting in prediction errors for systems that do not match the chosen analytical function.
On the other hand, non-parametric methods \cite{tan_control_2001,iyer_hysteresis_2004, ruderman_identification_2012} discretize the density function using a mesh grid (Fig.~\ref{fig:cartoon}b) and attempt to determine the density of hysterons at each mesh point based on experimental measurements.
However, these methods require large amounts of data or significant computational expense to fit high fidelity models.

We improve upon non-parametric modeling of hysteresis by creating \emph{differentiable} Preisach models, which use gradient-based optimization to identify the hysteron density function at high fidelities.
Differentiable modeling refers to tracking derivative information during every step of internal model calculations.
This allows what is known as back-propagation \cite{lecun_efficient_2012}, where through the chain rule, the derivative of the model output with respect to any model parameter is analytically calculable. 
By combining this technique with gradient based optimization algorithms (e.g. L-BFGS-B \cite{byrd_limited_1995} or Adam \cite{kingma_adam_2017}), we are able to scale non-parametric Preisach models to thousands of mesh points, while still being computationally cheap enough for use in online modeling.

We construct a differentiable Preisach model by implementing the non-parametric version of Eq.~\ref{eqn:preisach_model} in the python library \emph{PyTorch} \cite{paszke_pytorch_2019}.
The continuous hysteron density $\mu(\alpha, \beta)$ is replaced with a discrete one, located on a triangular mesh containing N mesh points on the Preisach plane $\mu_i = \mu(\alpha_i,\beta_i)$, where $i=1,\dots,N$.
We also replace the hysteron operator $\hat{\gamma}_{\alpha\beta}$ with a differentiable approximation $\hat{\zeta}_{\alpha\beta}$, enabling differentiability with respect to $u(t)$ as shown in Fig. \ref{fig:cartoon}(a) (see Supplemental Materials for the exact form). 
The differentiable, non-parametric Preisach model is given by
\begin{equation}
    f(t) = \sum_{i=1}^N \mu_i\hat{\zeta}_{\alpha\beta,i}u(t)
    \label{eqn:diff_model}
\end{equation}
where $\hat{\zeta}_{\alpha\beta,i}$ is the differentiable hysteron operator at the ith mesh point.
\begin{figure*}
    \includegraphics*[width=\linewidth]{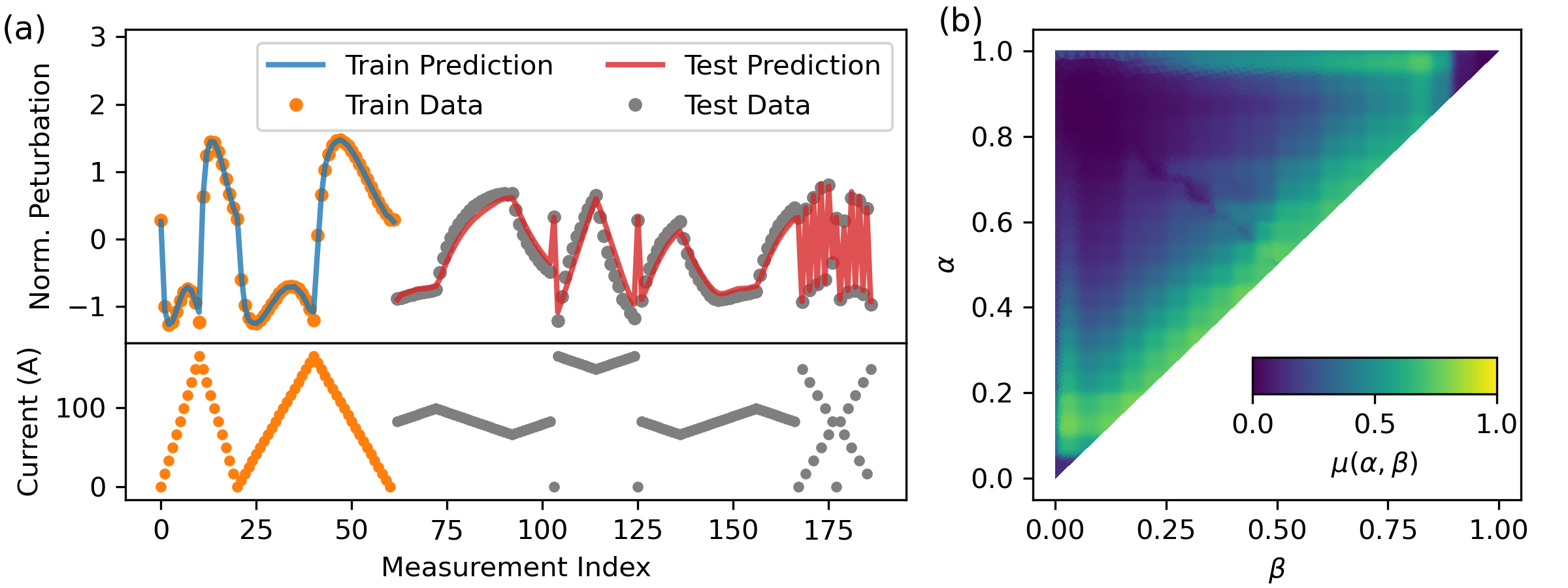}
    \caption{(Color online) Non-parametric modeling of hysteresis perturbations using direct measurements of a SLAC quadrupole magnet. (a) Normalized hysteresis error training/test data and model predictions. Sequence of applied currents during measurements $u(t)$ is normalized during training to the unit domain and measured field errors are transformed such that the training data has zero mean and a standard deviation of one. (b) Normalized hysteron density representing hysteresis perturbations $\bar{\mu}_i$ on the normalized Preisach plane after model training.}
   \label{fig:slac_fitting}
\end{figure*}

We demonstrate the effectiveness of our differentiable Preisach model by using it to analyze experimental data gathered from a SLAC quadrupole magnet.
Current applied to the magnet was cycled to sample both major and minor hysteresis loops and the integrated gradient at the magnet center was measured with a rotating coil measurement \cite{tanabe_iron_2005}.
Measurements were then split into training and test sets to investigate how accurately the model predicted measurement data and generalized to unknown future measurements.

Accelerator magnets pose a unique hysteresis modeling challenge, as they are designed specifically to minimize field perturbations from a polynomial function of magnet current due to hysteresis \cite{tanabe_iron_2005}.
As a result, we are are interested in resolving these field perturbations, which result from non-zero hysteron densities off of the $\alpha=\beta$ line, which we denote as $\bar{\mu}(\alpha,\beta)$.
Resolving these small perturbations requires specialized data processing and model construction, details of which can be found in the Supplemental Materials.

Model fitting to hysteresis perturbations observed in experiment is shown in Fig.~\ref{fig:slac_fitting} using an adaptive triangular mesh containing 7411 mesh points.
We trained the model on an Intel i9-9900K CPU at 3.6 GHz using a mean squared error loss function and the Adam algorithm with a learning rate of 0.01 over 10k steps, which took approximately 67~s. 
This is roughly two orders of magnitude faster than a comparable analysis in previous non-parametric studies \cite{iyer_hysteresis_2004} and could be improved further by limiting the number of optimization iterations, with minor degradation in model accuracy. 

Our model captures the features of major hysteresis loops with an RMS training error $\sigma_\text{train}$ of 0.8~mT, corresponding to a percentage error ($p=100\,\sigma_\text{train}/f_\text{max}$) of 0.015\%.
Despite only training on major hysteresis loops, our model makes accurate predictions of minor hysteresis loops and large swings in applied current with an RMS error of 2.6 mT ($0.051\%$).
Our model significantly outperforms polynomial fitting of the unnormalized experimental data, which has an RMS error of 12.1 mT ($0.23\%$) over the entire data set.

Next, we examine the case where \emph{directly measuring hysteresis output} is impractical or impossible.
For example, fields cannot be accurately characterized for magnetic elements that are already installed in accelerator beamlines.
Instead, we may only observe the beam response to fields generated by these elements.
To determine hysteresis characteristics in this case, we combine our hysteresis model with a Gaussian process (GP) model \cite{rasmussen_bayesian_2006} representing beam propagation as a function of magnetic fields. 
We then infer hysteresis behavior from measurements of the beam with respect to currents applied to each magnet.

\begin{figure}[h]
    \includegraphics[width=\linewidth]{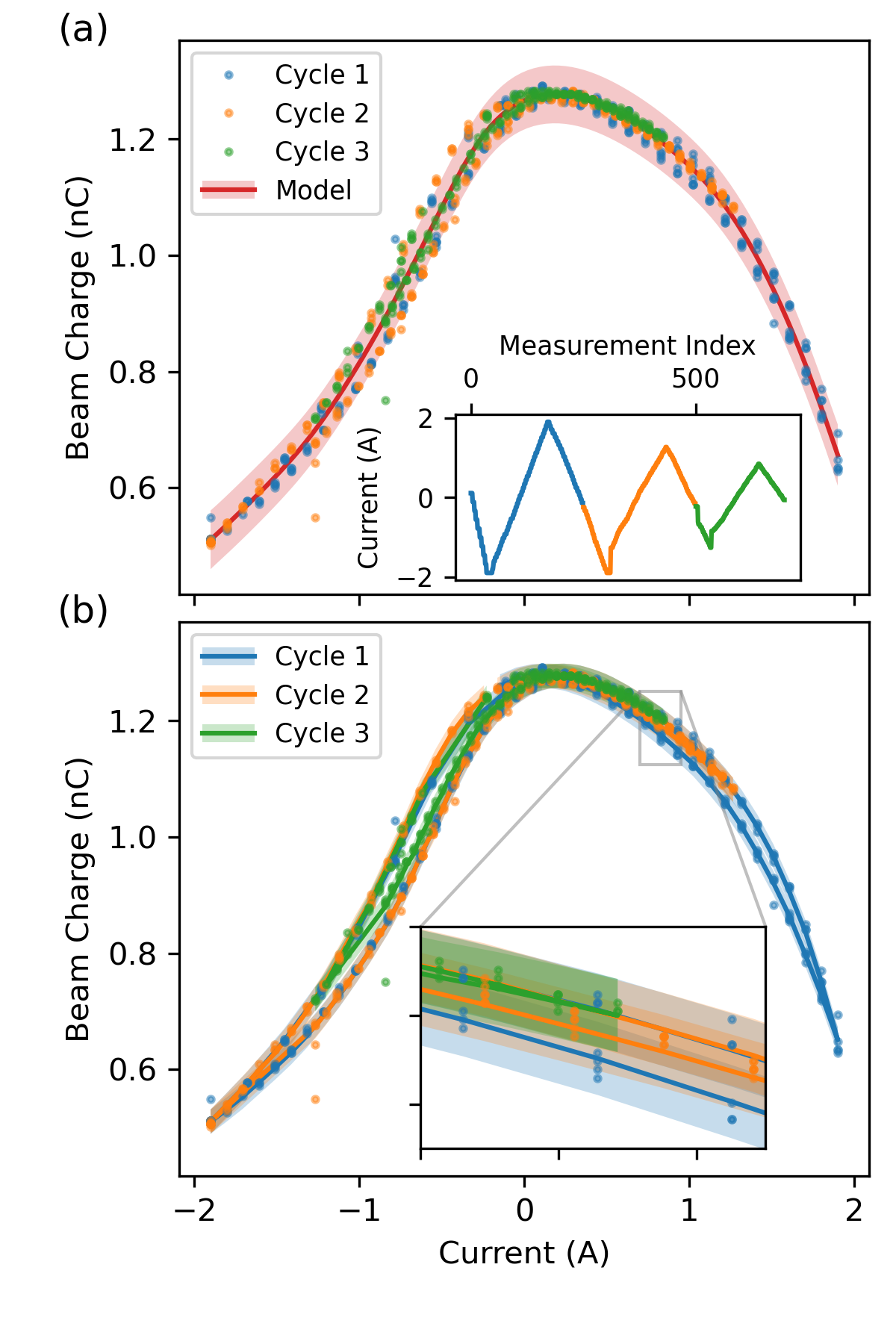}
    \caption{(Color online) Comparison between Gaussian process (GP) modeling and joint Hysteresis-GP modeling of beam transmission as a function of quadrupole current at the APS injector. (a) GP model prediction with training data over three cycles (see inset). Shading denotes 2$\sigma$ confidence region. (b) Hysteresis-GP model prediction, colored by cycle index.
    }
   \label{fig:aps_hybrid_fitting}
\end{figure}
The overall characteristics of GPs, defined as $g(\bm{x}) \sim \mathcal{GP}(m(\bm{x}), k(\bm{x},\bm{x}';\bm{\theta}))$ with a mean function $m(\bm{x})$ and covariance function $k(\bm{x},\bm{x}';\bm{\theta})$, are governed by a set of hyperparameters $\bm{\theta}$, which describe our prior knowledge of the model's smoothness, amplitude and noise.
GP models predict the distribution of function values at a location $\bm{x}$ to be $p(g | \mathcal{D}, \bm{x},\bm{\theta}) = \mathcal{N}(\mu(\bm{x}), \sigma^2(\bm{x}))$, where $\mathcal{D} = \{X,\bm{y}\}$ is the set of training samples and $\mu(\bm{x})$, $\sigma^2(\bm{x})$ are the posterior mean and uncertainty (see Supplemental Materials for details).
We infer hyperparameters for a GP model from training data by maximizing the marginal log likelihood given by
\begin{equation}
    p(\bm{y}|X) = \int p(\bm{y}|X,\bm{\theta})p(\bm{\theta}) d\bm{\theta}
    \label{eqn:mll}
\end{equation}
with respect to the hyperparameters $\bm{\theta}$, which results in a model that balances the trade-off between accuracy and complexity.

We combine the hysteresis and GP models into a single joint model by treating the hysteresis output as the GP input and training both models simultaneously.
The joint hysteresis-GP model is given by
\begin{equation}
    p(\bm{y}|\mathcal{D}, \bm{t}, \bm{\phi}, \bm{\theta}) = \mathcal{N}(\mu(f(\bm{t})), \sigma^2(f(\bm{t}))). 
\end{equation}
where $\bm{\phi}$ represents hysteresis model parameters.
The joint set of parameters $\Phi = \{\bm{\phi}, \bm{\theta}\}$ is then determined by maximizing the marginal likelihood using Eq.~\ref{eqn:mll} with respect to the new set of parameters $\bm{\Phi}$.

We demonstrate the effectiveness of our joint hysteresis-GP model by fitting the beam response with respect to the current applied to a focusing magnet located in the Advanced Photon Source (APS) injector \cite{sun_recent_2021}.
The current of the quadrupole magnet was varied using a sawtooth pattern from positive to negative 2 A while measuring beam charge passing through a downstream current monitor.
Measurements from this experiment, shown in Fig. \ref{fig:aps_hybrid_fitting}, have two sources of uncertainty, one from random noise inherent in the accelerator (aleatoric uncertainty) and one due to the unknown properties of magnetic hysteresis (epistemic uncertainty).
A normal GP model (Fig. \ref{fig:aps_hybrid_fitting}a) does not take into account the existence of hysteresis, thus it interprets epistemic errors due to hysteresis as aleatoric uncertainty, overestimating uncertainties in portions of the input domain. 
However, the joint hysteresis-GP model (Fig. \ref{fig:aps_hybrid_fitting}b), is able to resolve hysteresis cycles inside the data, removing epistemic uncertainties in the model prediction, thus improving model accuracy and reducing uncertainty.
The increase in accuracy from joint hysteresis-GP models has ramifications for model-based, online optimization of accelerators using BO, where we combine our online model of the accelerator with an acquisition function that chooses the next point to observe based on the model.

We examine how models with and without hysteresis taken into account affect optimization performance when optimizing a simulated accelerator which contains realistic magnetic elements that exhibit hysteresis.
We simulate the task of sequentially optimizing currents applied to 3 quadrupoles using BO, in order to transform an incoming round beam with an RMS beam size of $\sigma_{x,y} = 5$ mm to a final round beam size of $\sigma_\text{target} = 8$ mm.
The objective function is given by a geometric mean of the beamsize deviation $l=\sqrt{\Delta_x \Delta_y}$ where $\Delta_k = |\sigma_k - \sigma_\text{target}|$.
A toy hysteresis model (described in the Supplemental Materials) with a tunable hysteresis magnitude was used to simulate realistic magnetic elements.
Three beamlines with maximum fractional hysteresis errors $H_\epsilon=0,0.1,0.4$ were used in optimization trials to represent ideal, realistic, and extreme hysteresis effects respectively.

\begin{figure}[ht]
    \includegraphics*[width=\linewidth]{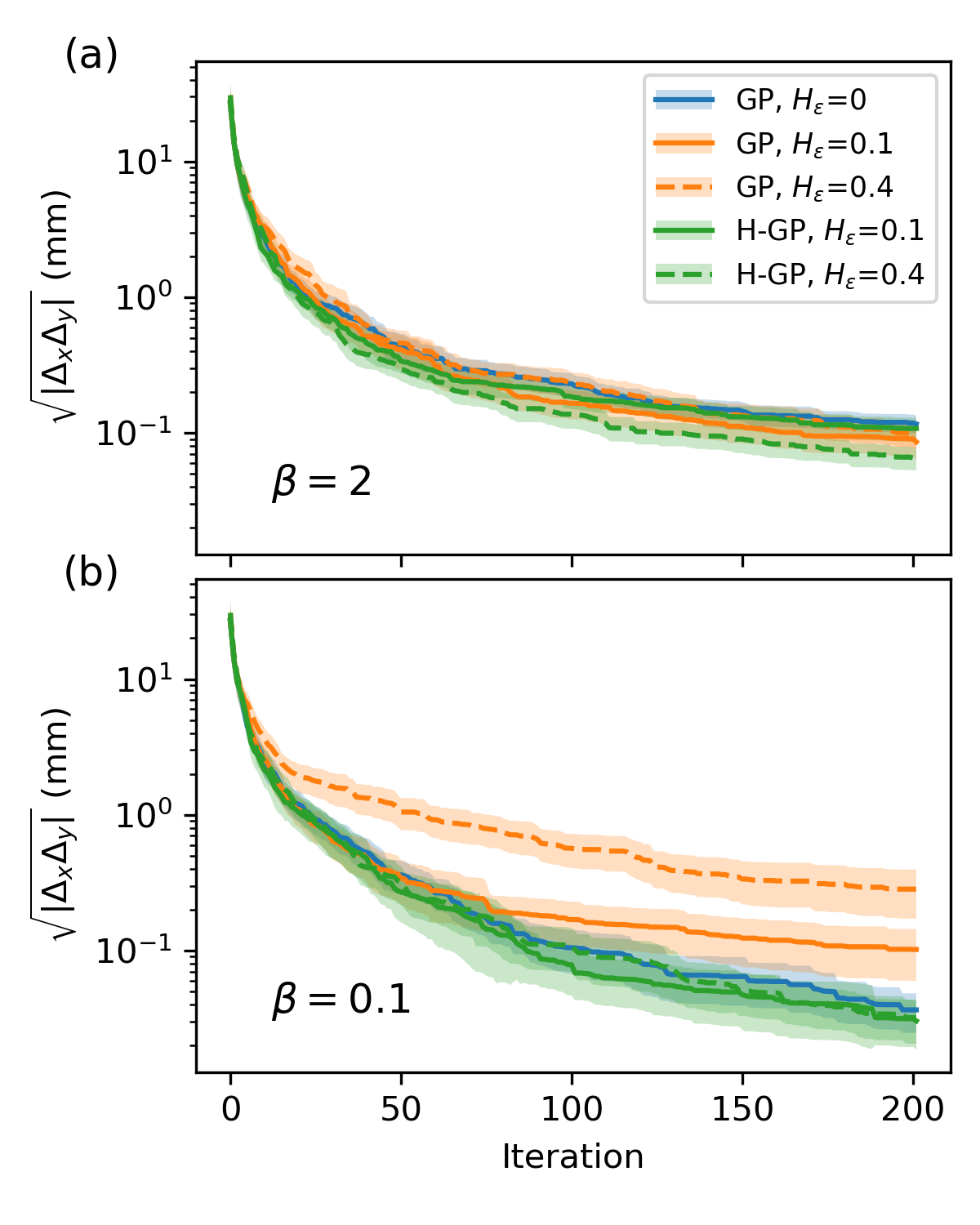}
    \caption{(Color online) BO performance for simulated beamline optimization problem using normal GP and joint hysteresis-GP (H-GP) models for varying maximum hysteresis errors $H_\epsilon$. (a) Performance when UCB is used with an even weighting of exploration and exploitation ($\beta=2$) and with strong weighting towards exploitation ($\beta=0.1$). Lines denote the mean best performance of 64 optimization runs, while shading denotes standard error.}
   \label{fig:optimization_comparison}
\end{figure}

We performed BO using the Upper Confidence Bound acquisition function \cite{srinivas_gaussian_2010}, first with $\beta=2$ which balances exploration (sampling points in unexplored regions of input space) and exploitation (sampling points that are predicted to be at global extrema).
We then repeated the experiment with $\beta=0.1$, which prioritizes exploitation.
Optimization results obtained over 64 trials using BO with normal GP models and joint hysteresis-GP (H-GP) models are shown in Figure \ref{fig:optimization_comparison}. 

Figure \ref{fig:optimization_comparison}a shows that hysteresis has little effect on the performance of BO when balancing exploration and exploitation, even when extreme hysteresis errors are present.
The acquisition function in this case often chooses to measure points in unexplored regions of input space and as a result, is relatively insensitive to hysteresis errors.

On the other hand, if we attempt to exploit the model as shown in Fig.~\ref{fig:optimization_comparison}b, modeling errors due to hysteresis effects in normal GP models negatively impact optimization performance, depending on the magnitude of hysteresis errors.
A joint hysteresis-GP model significantly improves optimization performance, matching the performance observed when optimizing an idealized beamline without hysteresis.

These results identify a clear strategy for optimizing physical systems that contain hysteresis using BO.
Hysteresis effects can be neglected when coarsely searching for global optima of a system, even when hysteresis errors are significant, since the optimization is dominated by uncertainties of unexplored regions in input space (especially in high dimensional input spaces where uncertainties are large).
However, hysteresis effects must be taken into account when attempting to exploit extrema during optimization.
Our model thus enables a staged approach toward optimization of systems involving hysteresis, where model and computational complexity is traded for optimization precision. 

In conclusion, we have demonstrated how a differentiable Preisach model can describe arbitrary hysteretic systems using direct or indirect measurements, and improve model based optimization of those systems.
Improved identification speed of differentiable Preisach modeling enabled practical, high-fidelity regression of major and minor hysteresis loops in realistic magnetic elements.
We were able to demonstrate that our hysteresis model can be combined with GP modeling to infer hysteresis behavior without making direct measurements of the hysteresis response.
Finally, we demonstrated how these joint hysteresis-GP models can be used to optimize physical systems containing hysteretic behavior, overcoming limitations faced by currently-used algorithms that neglect hysteresis repeatability errors.

Our work will support more advanced modeling of hysteresis in the future, most notably implementing fully Bayesian hysteresis models.
A fully Bayesian treatment of hysteresis involves replacing the hysteron density at each mesh point with a probability distribution and inferring the distribution using experimental data and Bayes' rule.
Differentiable modeling is required to implement this interpretation of the Preisach model through the use of stochastic variational inference, which turns the problem of probabilistic inference into an optimization one \cite{wingate_automated_2013}.
This would allow us to specify a prior distribution for the hysteron density at each mesh point, encoding explicit correlations between hysteron densities at different mesh points.
For example, we can strongly correlate hysteron densities between nearby mesh points, resulting in a smoother prediction of the overall hysteron density function.

\begin{acknowledgments}
The authors would like to thank the SLAC Metrology group for their help gathering quadrupole data and Louis Emery from the Advanced Photon Source for early discussions and preliminary data. This work was supported by the U.S. Department of Energy, under DOE Contract No. DE-AC02-76SF00515 and the Office of Science, Office of Basic Energy Sciences.

\end{acknowledgments}

\bibliography{prl_hysteresis}

\end{document}


\title{Supplemental Materials}

\maketitle
\section{Gaussian Process Modeling}
A GP represents the function value at a given input point via a random variable drawn from a joint Gaussian distribution $g(\mathbf{x}) \sim \mathcal{GP}(\mu(\mathbf{x}), k(\mathbf{x},\mathbf{x'}))$ where $\mu(\mathbf{x})$ is the mean and $k(\mathbf{x},\mathbf{x'})$ is known as the kernel \cite{rasmussen_gaussian_2006}, where $x,x'$ represents points in the input space.
To make a model prediction at the location $\bm{x}_*$ we start with a prior Gaussian distribution with $\mu(\mathbf{x})=0$ (without loss of generality) and the covariance matrix given by $k(\bm{x}_*,\bm{x}_*)$.
We then condition the joint multivariate Gaussian distribution on the data set of $N$ observed points $\mathcal{D}_N = \{X, \bm{y}\}$.
This gives us a probability distribution of the function value at the test point $g_* = g(\mathbf{x}_*)$ with expected noise $\sigma_n^2$ by
\begin{align}
    p(g_*|\mathcal{D}_{N}) &\sim \mathcal{N}(\mu_*,\sigma^2_*)\\
    \mu_* =&\ \mathbf{k}^T[K + \sigma_n^2 \mathbf{I}]^{-1}\mathbf{y}\\
    \sigma_* =&\ k(\mathbf{x}_*,\mathbf{x}_*) - \mathbf{k}^T[K + \sigma_n^2 \mathbf{I}]^{-1}\mathbf{k} \label{eqn:uncert}\\ 
    \mathbf{k} =&\ [k(\mathbf{x}_*,\mathbf{x}_1),k(\mathbf{x}_*,\mathbf{x}_1),\dots,k(\mathbf{x}_*,\mathbf{x}_N)]\\
    K =&\ 
    \begin{bmatrix}
        k(\mathbf{x}_1,\mathbf{x}_1) & \cdots & k(\mathbf{x}_1,\mathbf{x}_N)\\
        \vdots & \ddots & \vdots\\
        k(\mathbf{x}_N,\mathbf{x}_1) & \cdots & k(\mathbf{x}_N,\mathbf{x}_N)
    \end{bmatrix}
\end{align}
For this work we choose a Matern 5/2 kernel \cite{matern_spatial_2013}, based on prior knowledge of the target functions' smoothness.

\begin{figure}[ht]
    \includegraphics[width=\linewidth]{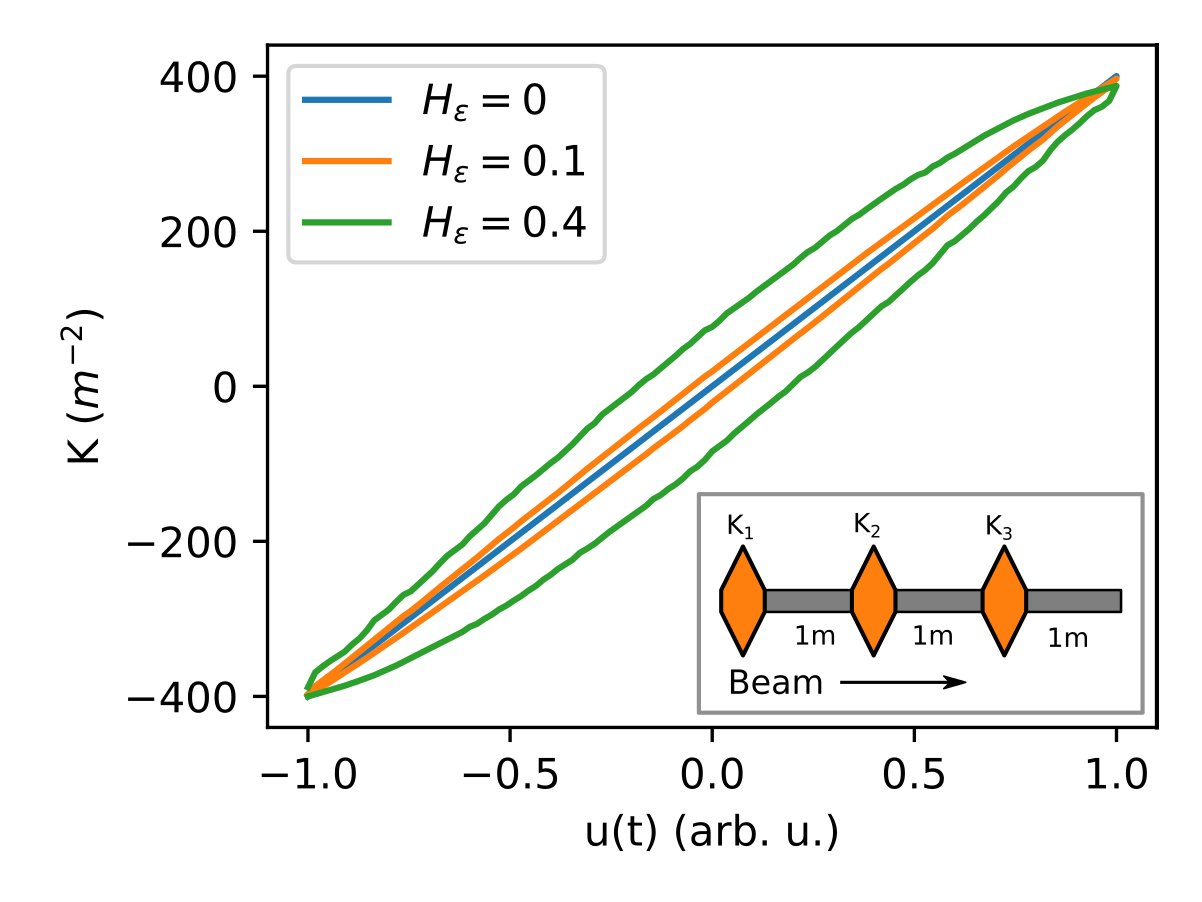}
    \caption{(Color online) Geometric quadrupole strengths as a function of input parameter $u(t)$ for quadrupoles used in optimization performance tests. Inset: Test beamline configuration (not to scale).}
    \label{fig:magnetization}
\end{figure}

\section{Differentiable Preisach Modeling for Accelerator Magnets}
Magnetic elements in accelerators are generally designed to reduce, but not eliminate completely, magnetic field errors due to hysteresis.
We start with a data set of inputs $u_m=u(t_m)$ and outputs $y_m=y(t_m)$ at given time steps $t_m, m=1\dots M$.
Since we wish to resolve small errors in experimental data $\mathcal{D} = \{u_m, y_m\}$ due to hysteresis, we construct the following model
\begin{equation}
    y_m = h(u_m) + \bar{f}(u_m) + \epsilon
\end{equation}
where $h(u_m)$ is a polynomial fit of the data using least squares fitting, $\bar{f}(u_m)$ is a Preisach model of the field errors, and $\epsilon$ is Gaussian noise.
The polynomial term is equivalent to a Preisach model with the hysteron density $\mu_{poly} = h(\alpha)\delta(\alpha - \beta)$ \cite{mayergoyz_mathematical_2003}. 

We can then determine $\bar{f}(u_m)$ and corresponding hysteron density $\bar{\mu}_i$ from measured field errors $\bar{y}_m = y_m - h(u_m)$ using a least squares approach.
For training purposes, field errors are shifted and scaled such that the training data has a mean of zero and a standard deviation of one.
The field errors are not monotonic with respect to $u_m$, which cannot be reconstructed using a Preisach model \cite{mayergoyz_mathematical_2003}.
We adapt our model for the field errors by adding a linear term, parameterized by \textit{B}, thus accounting for non-monotonic behavior.
Finally, a constant offset, \textit{C}, is included to allow for regression of minor hysteresis loops, resulting in the following model for hysteresis errors
\begin{equation}
    \bar{f}(u_m) = \frac{A}{N}\sum_{i=1}^N \bar{\mu}_i\hat{\zeta}_{\alpha\beta,i}u_m + Bu_m + C
    \label{eqn:model}
\end{equation}
where the set of free parameters is given by $\phi = \{A, B, C, \bar{\mu}_1,\dots,\bar{\mu}_N\}$.

We use an adaptive mesh density to reduce mesh points that are likely to be small, namely away from the line $\alpha=\beta$.
The mesh density is specified by the following function $\rho(\alpha,\beta) = \rho_0(|\alpha-\beta| + 0.05)$ where $\rho_0$ is a model hyperparameter. 
Meshes were generated on the normalized Preisach plane using the python library \textit{pygmsh} \cite{schlomer_pygmsh_nodate}.

For the model to be differentiable with respect to the input we replace the hysteron operator $\hat{\gamma}_{\alpha\beta}$ \cite{mayergoyz_mathematical_2003} with the differentiable hysteron operator $\hat{\zeta}_{\alpha\beta}$.
The output of this operator is given by
\begin{multline}
    \hat{\zeta}_{\alpha\beta}u_m = \\
    \begin{cases}
        \text{min}(\hat{\zeta}_{\alpha\beta}u_{m-1} + \tanh((u_m - \beta) / |T|),\ 1) & u_m > u_{m-1}\\
        \text{max}(\hat{\zeta}_{\alpha\beta}u_{m-1} - \tanh(-(u_m - \alpha) / |T|),\ -1) & u_m \leq u_{m-1}\\
    \end{cases}
\end{multline}
where $T$ is a temperature parameter that controls the maximum derivative of the operator. As $T\to 0$ this operator tends to $\hat{\gamma}_{\alpha\beta}$.
We use a value of $T=10^{-2}$ in the work done here.

The best fit free parameters $\phi^*$ are determined by minimizing the least squares loss function
\begin{equation}
    \phi^* = \textrm{argmin}_\phi\sum_{m=1}^M (\bar{y}_m - \bar{f}(u_m))^2
\end{equation}

\section{Optimization Problem Details}
We use a toy hysteresis model with a uniform hysteron density to simulate the optimization of quadrupole magnets with hysteresis.
The toy hysteresis model with a maximum field of $\pm1$ over the domain $u_m \in [-1,1]$ takes the following form
\begin{equation}
    f(u_m) = \frac{A}{N}\sum_{i=1}^N\hat{\zeta}_{\alpha\beta,i}u_m + (1 - A)u_m
\end{equation}
where $A$ is a scale parameter that controls the magnitude of hysteresis errors.

To investigate Bayesian optimization in the context of hysteresis we simulate a round electron beam passing through a beamline with three tunable quadrupoles, each with a thickness of 10 cm and geometric strength given by $K_i = 400[m^{-2}]f_i(t)$ where $i=1,2,3$.
Figure \ref{fig:magnetization} shows the geometric quadrupole strength as a function of $u(t)$ for the three toy hysteresis models used in optimization studies.
The beam starts with an initial transverse size of $\sigma_{x,y} = 5$ mm and divergences $\sigma_{x',y'} = 0.1$ mrad..
Beam propagation through the beamline (shown in Fig. \ref{fig:magnetization} is calculated to first order using drift and quadrupole thick-lens transport matrices \cite{weideman_particle_1999}.

\section{Effects from Training Data Sampling and Mesh Resolution on Hysteron Density}
\begin{figure*}[ht]
    \includegraphics[width=\linewidth]{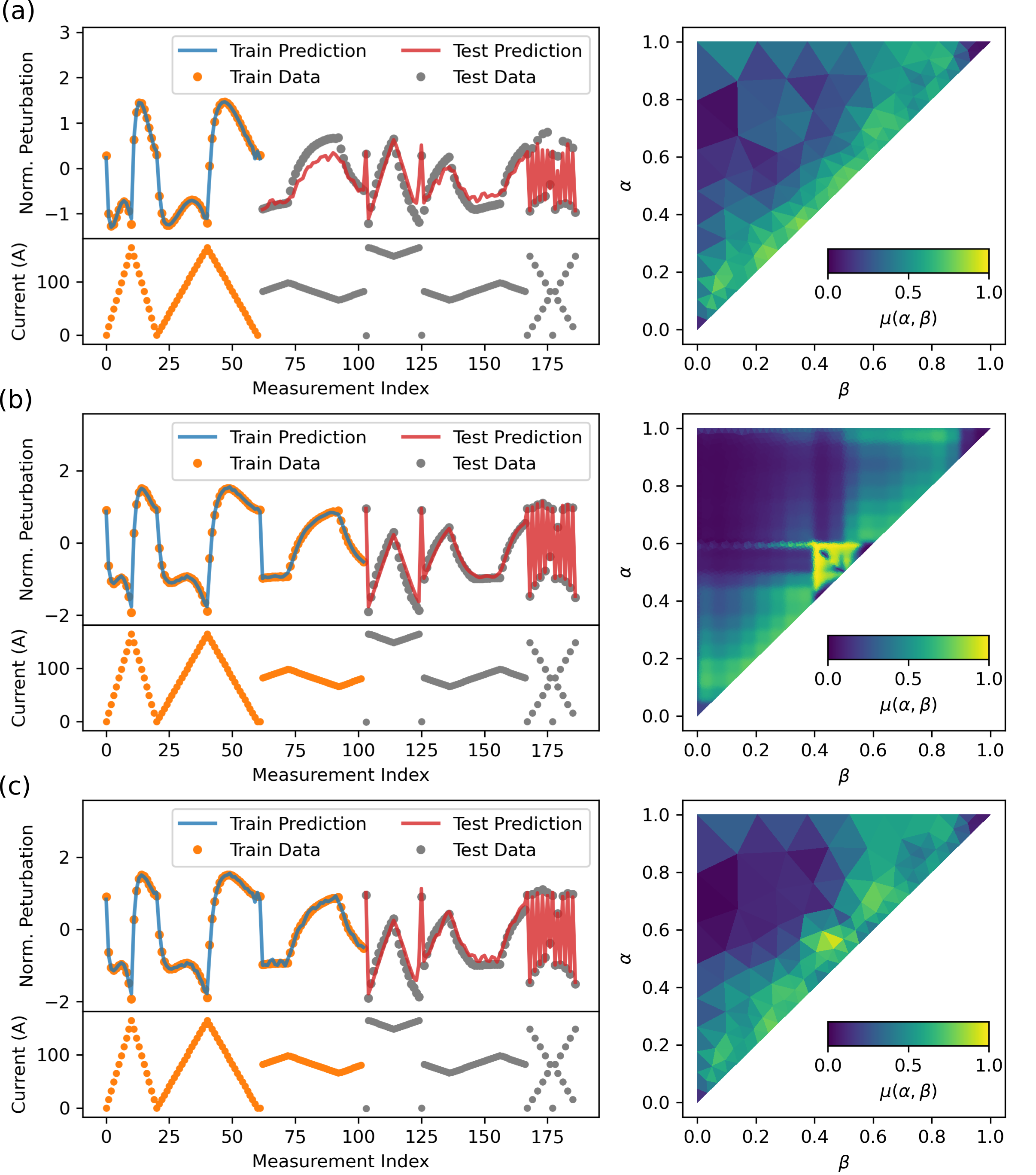}
    \caption{(Color online) Training results from varying training data sets based on measurements of SLAC quadrupole. Training data sets (left) and normalized error hysteron densities $\bar{\mu}_i$ (right) are plotted for (a) $N=62, r_{mesh}=0.05$, (b) $N=103, r_{mesh}=0.005$, and (c) $N=103, r_{mesh}=0.05$. Note that data is normalized based on the mean and spread of training points, resulting in different normalized values.}
   \label{fig:mesh_and_sampling_effects}
\end{figure*}

The predicted hysteron density $\bar{\mu}_i$ is strongly dependant on what points are sampled in the training data set and the resolution of the mesh.
Examples of these effects are shown in Fig. \ref{fig:mesh_and_sampling_effects}, where we refit the SLAC quadrupole data while modifying the mesh resolution and training data set.
Error characterization statistics between the model and experimental data for various training data sub-sets and mesh resolutions are seen in Table \ref{tab:errors}.
We observe that when a higher resolution mesh is used, both training and testing accuracy are improved, regardless of the number of training points.

\begin{table}[b]
\caption{\label{tab:errors} Training and testing errors of models for varying training/testing cutoff values $N_\text{train}$ and minimum normalized mesh resolution $r_\text{mesh}$.}
\begin{ruledtabular}
\begin{tabular}{llll}
\textrm{$N_\text{train}$}&
\textrm{$r_\text{mesh} [N]$}&
\textrm{$\sigma_\text{train}$ (mT)}&
\textrm{$\sigma_\text{test}$ (mT)}\\
\colrule
62 & 0.005 [7411] & 0.8 & 2.6\\
62 & 0.05 [111] & 0.9 & 3.9\\
103 & 0.005 [7411] & 0.7 & 2.2\\
103 & 0.05 [111] & 0.9 & 2.9\\
187 & 0.005 [7411] & 1.3 & N/A\\
187 & 0.05 [111] & 1.5 & N/A\\
\end{tabular}
\end{ruledtabular}
\end{table}

If the spacing between mesh points is significantly smaller than the spacing between measurements in the training data, the hysteron density can be underestimated as multiple mesh points are traversed in-between subsequent measurements, thus reducing the density fit for each mesh point.
As a result, mesh points in regions of the Preisach plane with lower measurement frequencies appear to have significantly lower densities, as shown in Fig. \ref{fig:mesh_and_sampling_effects}b.
However, when the mesh spacing increases. this effect disappears, as evident in Fig. \ref{fig:mesh_and_sampling_effects}c.
Regardless, our model still accurately captures the training data and generalizes well when making test predictions.

\section{Code and Data Availability}
Code and data used to produce figures in this manuscript are available in the public repository at \url{https://github.com/roussel-ryan/diff_hysteresis}.

\bibliography{prl_hysteresis}